\documentclass[12pt]{iopart}

\usepackage{iopams}

\usepackage{theorem}
\usepackage[dvipdfmx]{graphicx}
\usepackage[dvipdfmx]{hyperref}

\theoremstyle{break}

\newtheorem{Theorem}{Theorem}
\newtheorem{Proof}{Proof}

\newtheorem{Proposition}[Theorem]{Proposition}
\newtheorem{Corollary}[Theorem]{Corollary}
\newtheorem{Example}{Example}

\def\ds{\displaystyle}
\def\qed{\hfill\hbox{$\Box$}\vspace{0pt}\break}

\def\C{{\mathbb C}}

\def\R{{\mathbb R}}
\def\Z{{\mathbb Z}}

\def\1{\boldsymbol{1}}

\begin{document}

\title{Hankel Determinant Solution for Elliptic Sequence}

\author{Fumitaka Yura}

\address{
  Department of Complex and Intelligent Systems,
  Future University HAKODATE, 
  116-2 Kamedanakano-cho Hakodate Hokkaido, 041-8655, Japan
}
\ead{yura@fun.ac.jp}
\begin{abstract}
We show that the Hankel determinants of a generalized Catalan sequence satisfy
  the equations of the elliptic sequence. 
As a consequence, the coordinates of the multiples of an arbitrary point  
  on the elliptic curve are expressed by the Hankel determinants. 
\end{abstract}


\section{Introduction}
\label{sec:intro}

Our main purpose in this paper is to express the solution of the elliptic sequence
  by means of the Hankel determinants. 
The general solutions of the elliptic sequences and Somos 4 over $\C$ have been
  analytically studied by the elliptic function
  via the corresponding elliptic curves\cite{Ward1948, Ramani2002, Hone2005,Braden2005}. 
Meanwhile in the integrable systems, it is known that the Toda and Painlev\'e equations 
  have the Hankel determinant formula\cite{Kajiwara2001}. 
Therefore it is natural we presume the elliptic sequence also have 
  the Hankel determinant formulae. 
In section \ref{sec:ESS}, we briefly summarize the related preceding studies.
The main results and examples are shown in section \ref{sec:main},
  and in section \ref{sec:Somos4solution} the solution of Somos-(4)  
  is shown by our parametrization as an application, 
  which was obtained in \cite{Xin2009}. 
The last section \ref{sec:remarks} is devoted to the summary and
  the appendix describes the proof of the theorem.

\section{Elliptic sequences and Somos 4}
\label{sec:ESS}

\subsection{Elliptic sequence}
\label{sec:ES}
An {\it elliptic sequence} is defined by the equation
\begin{equation}
  W_{m+n} W_{m-n} = W_{m+1} W_{m-1} W_{n}^2 - W_{n+1} W_{n-1} W_{m}^2, 
  \quad m, n \in \Z .
\label{eq:EDSmn}
\end{equation}
It is easy to show $W_{-n} = -W_n$, and there is no loss of generality in taking $W_1=+1$. 
If $\left\{ W_i \right\}$ is an integer sequence and 
  $W_n$ divides $W_m$ whenever $n$ divides $m$, 
  the sequence $\left\{ W_i \right\}$ is called elliptic {\it divisibility} sequence (EDS).
Morgan Ward\cite{Ward1948} showed that $\left\{ W_i \right\}$ is an EDS if and only if 
  $W_2$, $W_3$ and $W_4$ are integers, and $W_2$ divides $W_4$.
He also showed that the elliptic sequence may be parametrized by 
  the Weierstrass sigma function $\sigma (z)$ as 
\begin{equation}
   W_n = \sigma (nz) /\sigma (z)^{n^2} 
\label{eq:sigma}
\end{equation}
  for the case $W_2 W_3 \neq 0$, from which the name {\it elliptic} comes. 
The sequence with this condition $W_2 W_3 \neq 0$ is called {\it proper}\cite{Ward1948}. 

In the case $n=2$, the equation (\ref{eq:EDSmn}) turns into
\begin{equation}
  W_{m+2} W_{m-2} = W_{2}^2 W_{m+1} W_{m-1} - W_{1} W_{3} W_{m}^2
  \quad (m \geq 3), 
\label{eq:EDS2}
\end{equation}
and
\begin{equation}
r_{n-1} r_{n}^2 r_{n+1} = W_{2}^2 r_{n} -W_1 W_3, 
  \label{eq:QRTr}
\end{equation}
with the definition 
\begin{equation}
  r_{n} := \frac{W_{n-1}W_{n+1}}{W_{n}^2} \quad (n \neq 0). 
\end{equation}
Note that this type of equation (\ref{eq:QRTr})
  is the special case of QRT mappings\cite{Ramani1995,Ramani2002}. 
The elliptic sequence $\left\{ W_n \right\}$ has one conserved quantity $I \equiv I_n$,
  which is independent of $n$: 
\begin{eqnarray}
  I_n & := & \frac{W_{n-2}W_{n+1}}{W_{n-1}W_{n}} 
           + \left( \frac{W_{2}}{W_{1}} \right)^2 \frac{W_{n}^2}{W_{n-1}W_{n+1}}
           + \frac{W_{n-1}W_{n+2}}{W_{n}W_{n+1}} 
\label{eq:cq} \\ 
  & = & r_{n-1}r_{n} + (W_{2}/W_{1})^2 /r_{n} + r_{n}r_{n+1}. \nonumber 
\end{eqnarray}

Let $E$ be an elliptic curve over a field $K$ that is given
  by Weierstrass form \cite{Shipsey2000, Swart2003, Silverman1986}
\begin{equation}
  y^2 + c_1 x y + c_3 y = x^3 + c_2 x^2 + c_4 x + c_6, 
  \label{eq:Weierstrass}
\end{equation}
and the constants $b_2 := c_1^2+4c_2$, $b_4 := c_1 c_3+2 c_4$, $b_6 := c_3^2+4 c_6$ and 
  $b_8 := c_1^2 c_6 - c_1 c_3 c_4 + 4 c_2 c_6 + c_2 c_3^2-c_4^2$. 
Let $P=(x_1, y_1)$ be a point on $E$. 
The {\it division polynomials} $\psi_n \equiv \psi_n(x_1, y_1)$ are defined by the following recursion:
\begin{eqnarray}
  \psi_0 & = & 0, \nonumber \\
  \psi_1 & = & 1, \nonumber \\
  \psi_2 & = & 2 y_1 + c_1 x_1 + c_3, \nonumber \\
  \psi_3 & = & 3 x_1^4 + b_2 x_1^3 + 3 b_4 x_1^2 + 3 b_6 x_1 + b_8, \nonumber \\
  \psi_4 & = & \psi_2 ( 2 x_1^6 + b_2 x_1^5 + 5 b_4 x_1^4 + 10 b_6 x_1^3 + 10 b_8 x_1^2  \label{eq:division} \\
           &  & \qquad  + (b_2 b_8-b_4 b_6) x_1 + b_4 b_8 - b_6^2), \nonumber \\
  \psi_{2n+1} & = & \psi_{n}^3 \psi_{n+2} - \psi_{n-1} \psi_{n+1}^3 \quad (n \geq 2), \nonumber \\
  \psi_{2n}  & = & (\psi_{n-1}^2 \psi_{n+2} - \psi_{n-2} \psi_{n+1}^2) \psi_{n}/\psi_{2} \quad (n \geq 3), \nonumber \\
  \psi_{-n}  & = & -\psi_{n} \quad (n < 0). \nonumber 
\end{eqnarray}
These polynomials are essentially the elliptic sequence, namely, $W_n \equiv \psi_n(x_1, y_1)$
  because the recursion relations (\ref{eq:division}) coincide with
  (\ref{eq:EDSmn}) in the case $m=n+2$ or $n+1$: 
\begin{equation}
\begin{array}{lcl}
  W_{2n} W_{2} & = & W_{n} (W_{n-1}^2 W_{n+2} - W_{n-2} W_{n+1}^2), \\
  W_{2n+1} & =&  W_{n}^3 W_{n+2} - W_{n-1} W_{n+1}^3. 
\end{array}
  \label{eq:Wdbl} 
\end{equation}

The coordinate $(x_n, y_n)$ of the point $nP := \overbrace{P + P + \cdots + P}^{n}$ on $E$ is expressed 
  by the division polynomials as
\begin{equation}
   nP = (x_n, y_n) := \left( 
  \frac{\theta_n (x_1, y_1)}{\psi_n (x_1, y_1)^2}, \frac{\omega_n (x_1, y_1)}{\psi_n (x_1, y_1)^3} \right), 
  \label{eq:nP}
\end{equation}
where $\theta_n(x_1, y_1) := x_1 \psi_n(x_1, y_1)^2 - \psi_{n-1}(x_1, y_1) \psi_{n+1}(x_1, y_1)$, 
and if $\mbox{char} (K) \neq 2$ and $n\neq 0$, 
$\ds \omega_n(x_1, y_1) := \frac{1}{2}\left(\frac{\psi_{2n}(x_1, y_1)}{\psi_{n}(x_1, y_1)}
  - \left(c_1 \theta_n(x_1, y_1) + c_3 \psi_{n}(x_1, y_1)^2 \right)  \psi_{n}(x_1, y_1) \right)$ \cite{Swart2003,Shipsey2000}. 
These relations are fundamentally of the elliptic functions  
\[
  \wp(nz) = \wp(z)-\phi_{n+1}(z)\phi_{n-1}(z)/\phi_{n}(z)^2, 
\]
where $\phi_n(z) = \sigma(nz)/\sigma(z)^{n^2}$. 
Let $Q=(q_{x}, q_{y})$ and $Q+nP = (\overline{x}_{n}, \overline{y}_{n})$ be also the points on $E$. 
The relations among $x$-coordinates $\overline{x}_{n}$, namely, $\wp$ functions,
  are presented in \cite{Hone2005,Poorten2006}, 
\begin{equation}
e_{n-1} e_{n}^2 e_{n+1} = \psi_2(x_1, y_1)^2 e_{n} - \psi_1(x_1, y_1) \psi_3(x_1, y_1) \label{eq:QRTe}, 
\end{equation}
where $e_n := x_1 - \overline{x}_{n}$. 
Note that (\ref{eq:QRTe}) is also the special case of QRT mappings,
  in which $e_n$ is shifted from $r_n$ by the {\it translation} $Q$\cite[sec.~4]{Poorten2006}. 
Let us finally define the sequence $\left\{ s_n \right\}_{n \geq 0}$
  by way of $s_{n-1} s_{n+1} = e_n s_n^2$ and initial values $s_0$, $s_1$. 
This transformation yields the Somos 4 equation (\ref{eq:somos4}) from (\ref{eq:QRTe}) 
  with $\alpha_1 = \psi_2(x_1, y_1)^2$ and $\alpha_2 = -\psi_1(x_1, y_1) \psi_3(x_1, y_1)$. 
In the next subsection, we will briefly sketch the Somos sequences. 

\subsection{Somos 4}
\label{sec:Somos4}

For $k \geq 4$, the Somos $k$ sequence $\left\{ s_i \right\}$ is defined by
\begin{equation}
  s_n s_{n-k} = \sum_{i=1}^{[k/2]} \alpha_i s_{n-i} s_{n-k+i} \quad (n\geq k). 
\label{eq:gensomosk}
\end{equation}
As the special case of the coefficients $\alpha_i =1 $ for all $i$ and 
  the initial values $s_0 = s_1 = \cdots = s_{k-1} = 1$, (\ref{eq:gensomosk}) gives
  the original Somos-($k$) sequence\cite{Somos1989, Gale1991}.
The surprising fact for $4 \leq k \leq 7$ is that the Somos-($k$) generates
  only integers $s_n$ for all $n$.
This {\it  integrality} is now understood as
  the Laurent property\cite{Fomin2002,Hone2008,Kanki2014a}. 
In this paper, we will consider only $k=4$ case; 
\begin{equation}
  s_{n-2} s_{n+2} =  \alpha_1 s_{n-1} s_{n+1} + \alpha_2 s_{n}^2 \quad (n\geq 2), 
\label{eq:somos4}
\end{equation}
where $\alpha_1$ and $\alpha_2$ are the constant coefficients,
  and $s_0, s_1, s_2$ and $s_3$ initial values. 
If we choose the six values $\alpha_1$, $\alpha_2$, $s_0$, $s_1$, $s_2$ and $s_3$,
  then $s_n$ for $n \geq 4$ are uniquely determined unless $s_{n-4} = 0$. 
The equation (\ref{eq:EDS2}) is apparently a special case of  (\ref{eq:somos4})
  with $\alpha_1 = W_2^2$, $\alpha_2 = -W_1 W_3$, $s_0=W_0$, $s_1=W_1$, $s_2=W_2$, $s_3=W_3$.
On the other hand, the result in section \ref{sec:ES} says that 
  (\ref{eq:somos4}) may be obtained from (\ref{eq:EDS2})
  if $\alpha_1$ is square or quadratic residue; 
  that is, (\ref{eq:somos4}) follows from the elliptic sequence
  with $W_2 = \pm \sqrt{\alpha_1}$, $W_3 = -\alpha_2$
  and the sequence $s_{n-1} s_{n+1} = e_n s_n^2$ with $e_n$ that is specified with
  $E$, $P$ and $Q$. 
The solutions of (\ref{eq:somos4}) is expressed as
\begin{equation}
  s_n = \frac{s_1^n}{s_0^{n-1}} e_1^{n-1} e_2^{n-2} \cdots e_{n-1} 
\label{eq:snbyen}
\end{equation}
by $e_1, e_2, \cdots, e_{n-1}$ and the initial values $s_0$ and $s_1$. 
In the paper\cite{Poorten2006}, the following identities
\begin{eqnarray}
  W_{m}^2 s_{n-t} s_{n+t} = W_{t}^2 s_{n-m} s_{n+m} - W_{t-m} W_{t+m} s_{n}^2, 
    \label{eq:comp1} \\
  W_{m} W_{m+1} s_{n-t} s_{n+t+1} = W_{t} W_{t+1} s_{n-m} s_{n+m+1} - W_{t-m} W_{t+m+1} s_{n} s_{n+1}, 
    \label{eq:comp2}
\end{eqnarray}
  were shown, as the title says ``Every Somos 4 is a Somos $k$'' for $k\geq 5$.  

Let us define the Hankel determinant $H^{(m)}_{n}$ for $m, n \geq 0$ of 
  the given sequence $\left\{ a_0, a_1, a_2, \ldots \right\}$ as
\begin{equation}
H_n^{(m)} := (a_{m+i+j})_{i, j=0}^{n-1} = 
\begin{array}{|cccc|}
a_m & a_{m+1} & \cdots & a_{m+n-1} \\
a_{m+1} & a_{m+2} & \cdots & a_{m+n} \\
\vdots & \vdots & \ddots & \vdots \\
a_{m+n-1} & a_{m+n} & \cdots & a_{m+2n-2}  
\end{array} \quad (n \geq 1),
\label{eq:defHankel}
\end{equation}
and the convention $H_0^{(m)} := 1$. 
The sequence of the Hankel determinants for $m=0$,  
  $\left\{ H^{(0)}_0, H^{(0)}_1, H^{(0)}_2, \ldots \right\}$, 
  is usually called the {\it Hankel transform} \cite{Layman2001} of $\left\{ a_n \right\}$. 
In \cite{Barry2010}, Paul Barry studied the families of generalized Catalan numbers
  with three parameters:
\begin{equation}
b_n = \left\{
\begin{array}{l}
1 \quad (n=0) \\
\alpha' \quad (n=1) \\
\alpha' b_{n-1}+\beta' b_{n-2}+\gamma' \sum_{i=0}^{n-2} b_i b_{n-2-i} \quad (n \geq 2) 
\end{array}
\right. , 
\label{eq:Barry}
\end{equation}
where $\alpha', \beta', \gamma'$ are constants. 
He conjectured that the Hankel transform of $\left\{ b_n \right\}$
  satisfies (\ref{eq:somos4}) by $s_n = H^{(0)}_n$, $\alpha_1 = \alpha'^2 \gamma'^2$,
  $\alpha_2 = \gamma'^2(\beta'+\gamma')^2-\alpha'^2 \gamma'^3$.
This conjecture was proved by Xiang-Ke Chang and Xing-Biao Hu\cite{Chang2012}. 
Note that Somos-($4$) seems not to be included in this parametrization. 
The original Somos-($4$)
  was solved in \cite{Xin2009}. 

\section{Solution of elliptic sequence by Hankel determinant}
\label{sec:main}

\subsection{Main theorem}
Let $a, b$ and $c$ be constants over a field $K$ and suppose
  the following sequence $\left\{ a_n \right\}$: 
\begin{equation}
  a_0 = a, a_1 = b, a_2 = c, a_{n+1} = \sum_{i=0}^{n} a_{i} a_{n-i} \ (n\geq 2),
  \label{eq:abc}
\end{equation} 
which is similar to $\left\{ b_n \right\}$ in (\ref{eq:Barry}). 
We refer to (\ref{eq:abc}) as the $(a, b, c)$-Catalan sequence. 
The so-called Catalan numbers may be retrieved from the (1, 1, 2)-Catalan sequence. 
By means of the Hankel matrices whose elements are the $(a, b, c)$-Catalan,
  we also define the sequence $\left\{ W_n \right\}_{n \in \Z}$, 
\begin{equation}
\left\{
\begin{array}{lcll}
W_{2n+1} & := & (-1)^n H^{(1)}_n & (n \geq 0)\\
W_{2n+2} & := & \sigma^n H^{(2)}_n W_2 & (n \geq 0) \\
W_{-n} & := & -W_{n} & (n \geq 0)
\end{array}
  \label{eq:defW}
\right. , 
\end{equation}
and impose one constraint  
 $W_2^4 = \sigma (2ab-c)$ with an arbitrary constant sign $\sigma = \pm 1$. 
For example, first few terms are calculated as  
\begin{equation}
\left\{
\begin{array}{lclcl}
& \vdots & & & \\
W_{-1} & = & -W_1 & = & -1 \\
W_{0} & = & 0 & & \\
W_{1} & = & H_0^{(1)} & = & 1 \\
W_{2} & & & & \\
W_{3} & = & -H_1^{(1)} & = & -b \\
W_{4} & = & \sigma H_1^{(2)} W_2 & = & \sigma c W_2 \\
W_{5} & = & H_2^{(1)} & = & b^3+2abc-c^2 \\
W_{6} & = & \sigma^2 H_2^{(2)} W_2 & = & -b (b^3+2abc-2c^2) W_2 \\
& \vdots & & & 
\end{array}
\right. 
\label{eq:exW}
\end{equation}
from $a_0 = a$, $a_1 = b$, $a_2 = c$, $a_3 = b^2+2ac$, $a_4 = 2(2a^2c+ab^2+bc)$,
  $a_5 = 4a^2 b^2+2b^3+8a^3c+8abc+c^2$, $\cdots$. 
Note that $W_2$ is not defined in (\ref{eq:defW}) since $\sigma^0=1$ and $H^{(2)}_0 = 1$. 
The following is our main theorem and the appendix is devoted to the proof: 
\begin{Theorem}
The double-sided infinite sequence $\left\{ W_n \right\}_{n \in \Z}$ defined above 
  satisfies the elliptic sequence (\ref{eq:EDS2}). 
\label{thm:main}
\end{Theorem}

This parametrization by $(a, b, c)$ of the elliptic sequence $\left\{ W_n \right\}$ is almost general: 
Let us consider (\ref{eq:defW}) over a field $K$ such that $\mbox{char} (K) \neq 2$. 
We may determine the $(a, b, c)$-Catalan sequence (\ref{eq:abc}) 
 from the four initial values $W_1(=1), W_2, W_3, W_4$ of the elliptic sequence by  
\begin{equation}
a = -\frac{\sigma}{2 W_3}\left( \frac{W_4}{W_2} + W_2^4 \right), 
  \quad b=-W_3, \quad c=\sigma \frac{W_4}{W_2}, 
\label{eq:abcFromW}
\end{equation}
as long as the sequence is proper $(W_2 W_3 \neq 0)$. 
Conversely, given $(a, b, c)$-Catalan sequence, the corresponding elliptic sequence should satisfy 
\[
W_2^4 = \sigma (2ab-c), \quad W_3 = -b, \quad W_4 = \sigma c W_2. 
\]
Provided that $\sigma (2ab-c)$ have the fourth root, the equation and its solution exist. 
If especially $K = \R$, choosing $\sigma$ as the same sign of $(2ab-c)$ always yields $W_2 = \sqrt[4]{\left| 2ab-c \right|}$. 
We note that the parameter $a$ in (\ref{eq:abcFromW}) is essentially $\wp'' (z)$
in the case $K = \C$ \cite{Ward1948}. 

For EDS, the parameter $a$ is not necessarily integer. 
The following is apparent from the above arguments and \cite{Ward1948}, 
  since $W_2$, $W_3$ and $W_4$ become integers and $W_2$ divides $W_4$: 
\begin{Corollary}
Let $W_2$, $b$ and $c$ be integers and $\sigma = \pm 1$ an arbitrary sign.
Then we always obtain EDS $\left\{ W_n \right\}$ by $(a, b, c)$-Catalan sequence,  
  where $a=(\sigma W_2^4+c)/(2b)$ if $b\neq 0$, otherwise $a$ is arbitrary ($b=0$).   
\end{Corollary}

\subsection{Curve and point sequence}
As shown in (\ref{eq:nP}), the elliptic sequence leads to the point sequence $\left\{ nP \right\}_{n}$. 
Hereafter we limit ourselves to the case of $\mbox{char}(K) \neq 2, 3$ and $W_2 \neq 0$ for simplicity. 
Suppose the elliptic sequence $\left\{ W_n \right\}$ by $(a, b, c)$-Catalan sequence. 
Then comparing (\ref{eq:exW}) with (\ref{eq:division}), we may express the point $P$
  on the curve $E: y^2 = x^3 + g_2 x + g_3$ associated with the elliptic sequence as 
\begin{eqnarray*}
  P & := & (x_1, y_1) =\left( \frac{a^2-b}{3W_2^2}, \frac{1}{2} W_2 \right), \\
  g_2 & = & -\frac{1}{3W_2^4} (a^4+4a^2 b+b^2-3ac) = -3x_1^2-\sigma a, \\
  g_3 & = & y_1^2 -x_1^3 -g_2 x_1 = 2 x_1^3+y_1^2+\sigma a x_1, 
\end{eqnarray*}
and therefore
\begin{equation}
   nP = (x_n, y_n) = \left( x_1 - \frac{W_{n-1}W_{n+1}}{W_{n}^2}, \frac{W_{2n}}{2 W_{n}^4} \right) \quad (n \neq 0)
\label{eq:nPbyW}
\end{equation}
from (\ref{eq:nP}). Solving these relations reversely yields the following:
\begin{Corollary}
Suppose the point $P=(x_1, y_1)$ on the curve $E: y^2 = x^3 + g_2 x + g_3$. 
Then the coordinates of $nP$ are given by
 (\ref{eq:nPbyW}) through $(a, b, c)$-Catalan sequence with 
\begin{eqnarray*}
  a & = & -\sigma(3 x_1^2 + g_2), \\
  b & = & -3 x_1^4-6g_2 x_1^2 - 12g_3 x_1 + g_2^2, \\
  c & = & 2 \sigma ( x_1^6+5 g_2 x_1^4 + 20 g_3 x_1^3 - 5g_2^2 x_1^2 - 4 g_2 g_3 x_1 - g_2^3 - 8 g_3^2 ),  \\
  W_2 & = & 2 y_1, 
\end{eqnarray*}
where $\sigma = \pm 1$ that is the common arbitrary sign. 
\end{Corollary}
We thus obtain the solution of the coordinates of $nP$ by means of the Hankel determinants. 
The conserved quantity in (\ref{eq:cq}) are also obtained as $I = -2\sigma a$
  from the initial conditions of the elliptic sequence $\left\{ W_n \right\}$.

\subsection{Examples}
In this subsection, we show some examples of parametrization from typical elliptic sequences. 

\begin{Example}
Suppose the solution $W_n = n$ of (\ref{eq:EDSmn}), which seems to be simplest 
  as in \cite{Ward1948}.
Notwithstanding, our representation becomes more complex than it looks. 
The corresponding parameters are given as $a=-3 \sigma$, $b=-3$, $c=2 \sigma, \sigma = \pm 1$
  and $W_2 = 2$. Simple calculations show
\begin{equation}
  \left\{ a_n \right\}_{n=0}^{\infty} = \left\{ -3\sigma, -3, 2 \sigma, -3, 6\sigma, -14, 36 \sigma, -99, \cdots \right\} ,
  \label{eq:Wn}
\end{equation}
and
\begin{eqnarray*}
  H^{(1)}_0 = 1, H^{(1)}_1 = -3, H^{(1)}_2 = 5, \cdots,  H^{(1)}_n = (-1)^n (2n+1), \cdots, \\ 
  H^{(2)}_1 = 2 \sigma, H^{(2)}_2 = 3, H^{(2)}_3 = 4\sigma, \cdots,  H^{(2)}_n = \sigma^n (n+1), \cdots .
\end{eqnarray*}
The sequence (\ref{eq:Wn}) with $\sigma = +1$ for $n \geq 1$ is
\href{http://oeis.org/A184881/}{A184881} in \cite{OEIS}; 
\[
    {}_2 F_{1} \left[ \begin{array}{c} -2n, -2n \\ 1 \end{array} ; -1 \right]
  - {}_2 F_{1} \left[ \begin{array}{c} -2n-2, -2n+2 \\ 1 \end{array} ; -1 \right], 
\]
  where ${}_p F_{q}$ is the hypergeometric series. 
We note that this example is the singular case $y^2=x^3$.  
\end{Example}

\begin{Example}[(1, 1, 2)-Catalan sequence]
\label{ex:Catalan}
The parameters $a=1$, $b=1$, $c=2$ gives $W_2=0$ and the Catalan sequence
  $a_n = C_n := \ds \frac{1}{n+1} \left( \begin{array}{c} 2n \\ n \end{array} \right)$, 
  which leads to 
\begin{equation*}
  W_n = \left\{ 
  \begin{array}{cl}
    1 & (n \equiv 1 \bmod 4) \\
    -1 & (n \equiv 3 \bmod 4) \\
    0 & (\mbox{otherwise})
  \end{array}
  \right. \quad \mbox{for all } n. 
\label{eq:exW2_0}
\end{equation*}
This example follows from the well-known facts that
  the Hankel determinants of the Catalan numbers are  
  $H^{(1)}_n = 1$ and $H^{(2)}_n = n+1$ for $n \geq 0$ $(\mbox{cf.}~(\ref{eq:TBS2rr}))$. 
\end{Example}

\begin{Example}[Fibonacci sequence]
The parameters $a = -\sigma/2, b=2, c=-3\sigma, W_2=1, \sigma = \pm 1$ yield 
\begin{equation*}
\left\{a_n \right\}_{n=0}^{\infty} = \left\{
  -\sigma/2, 2, -3\sigma, 7, -19\sigma, 56, -174\sigma, 561, \cdots
\right\}
\end{equation*}
and 
\[
  W_n = (-1)^{(n-1)(n-2)/2} F_n, \quad F_n=F_{n-1}+F_{n-2}, \quad F_{0}=0, \quad F_{1}=1, 
\]
  which is the Fibonacci sequence with signs. 
For the case of $\sigma = -1$, 
\[
\left\{a_n \right\}_{n=0}^{\infty} = \left\{
  1/2, 2, 3, 7, 19, 56, 174, 561, \cdots
\right\}
\]
is the sum of adjacent Catalan numbers\cite{Cvetkovic2002} and 
  \href{http://oeis.org/A005807/}{A005807} in \cite{OEIS}; 
that is, $a_n=C_{n-1}+C_{n}$ with the convention $C_{-1}=-1/2$ that satisfies
the relation of Catalan numbers $(4n+2)C_{n} = (n+2)C_{n+1}$ to negative direction. 
\end{Example}

\begin{Example}[Integer factorization]
Let us consider Lenstra's elliptic curve method (ECM)
  to find a factor of $N=5429=61\times 89$ (example in \cite{Koblitz1994}).
Suppose,  for example, the elliptic curve $E: y^2 = x^3+2x-2$ over $\Z_N$,
  and let $P$ be the point $(1, 1)$ on $E$
  that are randomly chosen in ECM algorithm. 
We then numerically obtain 
\[
2P=(4076, 3384), \cdots, 36P=(97, 2928), \cdots, 72P={\cal O}, 
\]
where ${\cal O}$ is the point at infinity. 
In this case, ECM finds an non-invertible element in the $y$-coordinate of $36P$ 
  due to the fail in addition-formula for $36P+36P$. Thus we obtain one of the prime factors 
$\gcd(2928, N) = 61$.

The parameters corresponding to the curve $E$ and the point $P$ above are determined as 
\[
a = -5 \sigma, \quad b = 13, \quad c = -146 \sigma, \quad W_2 = 2, \quad \sigma = \pm 1
\]
from (\ref{eq:division}). By these parameters, numerical calculation indeed yields  
$\gcd(H^{(2)}_{35}, N) = 61$ because the denominator of $72P$ is 
$\psi_{72}(P) \equiv W_{72} = \sigma W_2 H^{(2)}_{35}$ from (\ref{eq:nP}) and (\ref{eq:defW}). 
Note that the naive calculation of determinant is typically of
  cubic-order of the matrix size. 
There is no significance for application as is,
  compared with the addition-and-duplication method\cite[section 3.4]{Shipsey2000} by 
  (\ref{eq:Wdbl}). 
\end{Example}

\subsection{Equivalent sequences}
Two sequences $\left\{ W_n \right\}$ and $\left\{ \bar{W}_n \right\}$ are said to be {\it equivalent} 
  if and only if there exists a constant $\theta \neq 0$ such that
  $\bar{W}_n = \theta^{n^2-1} W_n$\cite{Ward1948}. 
Suppose the transformation $\bar{a}_i = \theta^{i+1} a_i$. 
This transformation leads to 
\begin{equation*}
  \bar{a}_0 = \theta a, \quad
  \bar{a}_1 = \theta^2 b, \quad
  \bar{a}_2 = \theta^3 c, \quad 
  \bar{a}_{n+1} = \sum_{i=0}^{n} \bar{a}_{i} \bar{a}_{n-i} \ (n\geq 2), 
\end{equation*} 
  namely, $(\theta a, \theta^2 b, \theta^3 c)$-Catalan sequence. 
Let the corresponding Hankel determinants and the elliptic sequence by
  $(\theta a, \theta ^2 b, \theta ^3 c)$-Catalan sequence be
  $\bar{H}^{(m)}_n$ and $\left\{ \bar{W}_n \right\}$, respectively. 
We then obtain 
\[
  \bar{H}^{(m)}_n = \theta^{n(n+m)} H^{(m)}_n. 
\]
If $n$ is odd, taking $n=2k+1$ yields
  $\bar{W}_n = \bar{W}_{2k+1} = (-1)^k \bar{H}^{(1)}_{k}
   = (-1)^k \theta^{k(k+1)} H^{(1)}_{k} = (\theta^{1/4})^{n^2-1} W_n $. 
Otherwise $n$ is even, $n=2k+2$ yields
  $\bar{W}_n = \bar{W}_{2k+2} = \sigma^k \bar{W}_2 \bar{H}^{(2)}_{k}
   = \sigma^k (\theta^{3/4} W_2) (\theta^{k(k+2)} H^{(2)}_{k}) = (\theta^{1/4})^{n^2-1} W_n $. 
Thus $(\theta a, \theta ^2 b, \theta ^3 c)$-Catalan sequences
  for all $\theta$($\neq 0$) are equivalent to $(a, b, c)$-Catalan sequence
 if $\theta^{1/4}$ exists over $K$.

\section{Solution for Somos-(4)}
\label{sec:Somos4solution}

The solution of the original Somos-($4$), which is the case $\alpha_1=\alpha_2=1$ in (\ref{eq:somos4}), 
  was obtained in \cite{Xin2009}. 
The aim of this section is to express the solution via our parametrization from an elliptic sequence. 
The first few terms of
\begin{equation}
  s_{n-2} s_{n+2} = s_{n-1} s_{n+1} + s_{n}^2 \quad (n\geq 2), \quad s_0 = s_1 = s_2 = s_3 = 1. 
\label{eq:somos4withivs}
\end{equation}
are calculated as 
\[
s_0 = s_1 = s_2 = s_3 = 1, s_4=2, s_5=3, s_6=7, s_7=23, s_8=59, s_9=314, \cdots, 
\]
and the sequence $\left\{ e_n \right\}$ is determined as 
\[
e_1 = 1, e_2=1, e_3=2, e_4=3/4, e_5=14/9, e_6=69/49, e_7=413/529, \cdots, 
\]
from $e_n = s_{n-1} s_{n+1}/s_{n}^2$. 
Choosing $m=1$ and $t=2$ in (\ref{eq:comp1}) leads to 
\begin{eqnarray*}
   W_{1}^2 s_{n-2} s_{n+2} & = & W_{2}^2 s_{n-1} s_{n+1} - W_{1} W_{3} s_{n}^2, 
\end{eqnarray*}
and results in $W_2^2=1$ and $b=-W_3=1$. 
Choosing $m=1$, $t=3$ and $n=3$ in (\ref{eq:comp1}) also leads to 
\begin{eqnarray*}
   W_{1}^2 s_{0} s_{6} & = & W_{3}^2 s_{2} s_{4} - W_{2} W_{4} s_{3}^2, 
\end{eqnarray*}
and $a=-2\sigma$, $c=-5\sigma$ via $W_4 = \sigma c W_2$. 
Therefore these parametrizations yield  
\begin{eqnarray*}
  W_{n-2}W_{n+2} = W_{n-1}W_{n+1} + W_{n}^2, \\
  W_0 = 0, W_1 = 1, W_2^2 = 1, W_3 = -1, W_4 = -5W_2, W_5 = -4, W_6 = 29W_2, \\
  \qquad W_7 = 129, W_8 = -65W_2, W_9 = -3689, W_{10} = -16264W_2, \cdots, \\
I = -2\sigma a = 4, 
\end{eqnarray*}
where each $W_n \ (n \neq 2)$ is already determined
  by not numerical way but the Hankel determinant
  through $(-2\sigma, 1, -5\sigma)$-Catalan sequence at this stage. We also obtain 
\begin{eqnarray*}
  r_1=0, r_2 = -1, r_3 = -5, r_4 = 4/25, r_5 = -145/16, r_6 = -516/841, \cdots, 
\end{eqnarray*}
where $r_n = W_{n-1}W_{n+1}/W_{n}^2$, and the elliptic curve $E$ and the point $P$ on $E$:
\begin{eqnarray*}
  E: y^2=x^3+g_2 x + g_3, \quad g_2 = -1, g_3 = 1/4, \\
  P = (x_1, y_1) =\left( 1, W_2/2 \right). 
\end{eqnarray*}

Next we solve the translation $Q$ from $Q+P=(\bar{x}_1, \bar{y}_1)$ and
  $Q+2P=(\bar{x}_2, \bar{y}_2)$. 
Because the $x$-coordinates are calculated as
  $\bar{x}_1 = x_1 - e_1 = 0$ and $\bar{x}_2 = x_1 - e_2 = 0$
  from $x_1 = 1$ and $e_1=e_2=1$,  
we may obtain
\begin{equation}
  2Q+3P={\cal O}
\label{eq:2Q3PO}
\end{equation}
 and $Q=(q_{x}, q_{y})=(-1, W_2/2)$ due to $P \neq {\cal O}$.
Note that this relation $Q+(Q+3P)={\cal O}$ follows from the fact that
  Somos-(4) sequence is even with respect to $n \leftrightarrow 3-n$, namely, $s_{3-n} = s_{n}$.  

By these parametrizations, Somos-(4) may be solved. 
The $x$-coordinate $\bar{x}_n$ of $Q+nP$ is calculated by addition formula, 
\begin{eqnarray*}
\bar{x}_n & = & \lambda^2 - q_{x} - x_n, \quad \lambda = \frac{q_{y}-y_n}{q_{x}-x_n}, 
\end{eqnarray*}
due to $q_{x} \neq x_n$, namely, $nP \pm Q \neq {\cal O}$. 
If not, $nP \pm Q = {\cal O}$ gives $(2n \pm 3)P = {\cal O}$,
  and this contradicts that the point $P$ is of infinite order, which follows from 
  the Nagell-Lutz theorem\cite{Silverman1992} 
  with the fact that the coordinates $(4x_n, 8y_n)$ contain non-integers. 
From lengthy calculations, we may obtain
\begin{eqnarray*}
e_n & = & x_1 - \bar{x}_n \\
  & = & \left( (r_{n+1}+2)r_{n}^2 - 8 r_n + 4 \right)/(2-r_n)^2, 
\end{eqnarray*}
and furthermore, 
\begin{eqnarray*}
e_n e_{n+3} & = & f_{n-1}f_{n+1}/f_{n}^2, 
\end{eqnarray*}
where we define $f_n := 2 W_n^2 - W_{n-1}W_{n+1}$. 
These relations with (\ref{eq:snbyen}) yield 
\begin{eqnarray}
\hspace*{-10mm} 
s_n s_{n+3} & = & 
\left( \frac{s_1^n}{s_0^{n-1}}      e_1^{n-1} e_2^{n-2} \cdots e_{n-1} \right)
\left( \frac{s_1^{n+3}}{s_0^{n+2}} e_1^{n+2} e_2^{n+1} \cdots e_{n+2}  \right) \nonumber \\
& = & 
\frac{s_1^{2n+3}}{s_0^{2n+1}} e_1^{n+2}e_2^{n+1}e_3^{n}
   (e_1 e_4)^{n-1} (e_2 e_5)^{n-2} \cdots (e_{n-2} e_{n+1})^2(e_{n-1} e_{n+2})^1 \nonumber \\
& = & 
\frac{s_1^{2n+3}}{s_0^{2n+1}} e_1^{n+2}e_2^{n+1}e_3^{n}
  \left(\frac{f_{0}f_{2}}{f_{1}^2}\right)^{n-1} 
  \left(\frac{f_{1}f_{3}}{f_{2}^2}\right)^{n-2}
  \left(\frac{f_{2}f_{4}}{f_{3}^2}\right)^{n-3}
    \cdots  \left(\frac{f_{n-2}f_{n}}{f_{n-1}^2}\right)^{1} \nonumber \\
& = & 
\frac{s_1^{2n+3}}{s_0^{2n+1}} e_1^{n+2}e_2^{n+1}e_3^{n} \times 
  \frac{f_{0}^{n-1}}{f_{1}^{n}} f_{n} \nonumber \\
& = & 
  f_{n}, \label{eq:ssf}
\end{eqnarray}
where the last equality follows from the constants
  $s_0 = s_1 = 1$, $e_1=1$, $e_2=1$, $e_3=2$,
  $f_0=2W_0^2-W_{-1}W_{1}=1$, $f_1=2W_1^2-W_0W_2=2$. 
Solving (\ref{eq:ssf}), we obtain the following formula:
\begin{equation}
s_{6m+k} = \frac{f_{6m+k-3}f_{6m+k-9} \cdots f_{k+3}}{f_{6m+k-6}f_{6m+k-12} \cdots f_{k}} s_{k}, 
\end{equation}
where $m \ge 1$ and $0 \le k \leq 5$, 
  and each $f_{n}$ is given by (\ref{eq:defW}) as
\begin{equation}
\begin{array}{lcl}
f_{2n} & = & 
  2W_{2n}^2-W_{2n-1}W_{2n+1} \\
  & = & 2 \left( H^{(2)}_{n-1} \right)^2 + H^{(1)}_{n-1} H^{(1)}_{n}, \\
f_{2n+1} & = & 
  2W_{2n+1}^2-W_{2n}W_{2n+2} \\
  & = & 2 \left( H^{(1)}_{n \vphantom{n+1} } \right)^2 - \sigma H^{(2)}_{n-1} H^{(2)}_{n}. 
\end{array}
\label{eq:fbyH}
\end{equation}
Note that not only Somos-(4) but Somos $4$ have solutions of this type. 

The above relation (\ref{eq:ssf}) recursively defines $s_{k}$. 
In Somos-($4$) case, we may obtain simpler form by means of (\ref{eq:2Q3PO}). 
Let $R$ be the point $(0, W_2/2)$,  
  then (\ref{eq:2Q3PO}) yields $P=2R$, $Q=-3R$ and $Q+nP=(2n-3)R$. 
Note that $Q+nP$ are generated by only $R$. 
This special property of Somos-($4$) leads to the following:
Suppose 
\begin{eqnarray*}
  E: y^2=x^3+g_2 x + g_3, \quad g_2 = -1, \quad g_3 = 1/4, \\
  R =\left( 0, W_2/2 \right), 
\end{eqnarray*}
then the corresponding $\left\{ nR \right\}_n$
  is given by the elliptic sequence $\hat{W}_n$ through
\begin{eqnarray*}
a=\sigma, \quad b=1, \quad c=\sigma, \quad \hat{W}_2^2=1, \quad \sigma = \pm 1 \\
\hat{W}_{n-2}\hat{W}_{n+2} = \hat{W}_{n-1}\hat{W}_{n-1} + \hat{W}_{n}^2, \quad
\hat{W}_1 = 1, \quad \hat{W}_3 = -1, \quad \hat{W}_4 = \hat{W}_2.  
\end{eqnarray*}
We obtain the $x$-coordinate $\bar{x}_n$ of $Q+nP$ as 
  $\bar{x}_n = - \hat{W}_{2n-4}\hat{W}_{2n-2}/\hat{W}_{2n-3}^2$. 
This yields another formula for $e_n$ as
\begin{eqnarray*}
e_n & = & x_1 - \bar{x}_n \\
& = & 1 + \hat{W}_{2n-4}\hat{W}_{2n-2}/\hat{W}_{2n-3}^2 \\
& = & \hat{W}_{2n-5}\hat{W}_{2n-1}/\hat{W}_{2n-3}^2, 
\end{eqnarray*}
and the dependent variables $s_n$ as 
\begin{eqnarray}
s_n & = & 
  \frac{s_1^n}{s_0^{n-1}} e_1^{n-1} e_2^{n-2} \cdots e_{n-1} \nonumber \\
& = & 
\frac{s_1^{n}}{s_0^{n-1}} e_1^{n-1}e_2^{n-2}
  \left(\frac{ \hat{W}_{1}\hat{W}_{5}}{\hat{W}_{3}^2}\right)^{n-1} 
  \left(\frac{ \hat{W}_{3}\hat{W}_{7}}{\hat{W}_{5}^2}\right)^{n-2} 
\cdots
    \left(\frac{ \hat{W}_{2n-7}\hat{W}_{2n-3}}{\hat{W}_{2n-5}^2}\right)^{1} 
 \nonumber \\
& = & 
\frac{s_1^{n}}{s_0^{n-1}} e_1^{n-1}e_2^{n-2}
  \left(\frac{ \hat{W}_{1}^{n-3}}{\hat{W}_{3}^{n-2}}\right) \hat{W}_{2n-3} 
 \nonumber \\
& = & 
  \hat{H}^{(1)}_{n-2},
\label{eq:sbyH}
\end{eqnarray}
where the last equality follows from the constants
  $s_0 = s_1 = 1$, $e_1=1$, $e_2=1$, 
  $W_1=1$, $W_3=-1$ and $W_{2n-3} = (-1)^{n-2} \hat{H}^{(1)}_{n-2}$. 
Here the matrix elements in the Hankel determinant $\hat{H}^{(1)}_{n-2}$ 
  are $(\sigma, 1, \sigma)$-Catalan numbers. 
As a result, the solution of Somos-($4$) is expressed by the single Hankel determinant, 
which coincides with \cite{Xin2009, Barry2012}. 

\begin{Example}[$n=6$]
Let us verify the above argument in the case $n=6$ as an example. 
We first calculate the $(a, b, c)$-Catalan sequence by (\ref{eq:abc}); 
$a_0 = a = -2 \sigma$, $a_1 = b = 1$, $a_2 = c = -5\sigma$, 
$a_3 = 2 a_0 a_2 + a_1^2 = 21$, $a_4 = 2(a_0 a_3 + a_1 a_2) = -94\sigma$, $a_5=443$, $\cdots$. 
Note that this sequence corresponds to $\left\{ nP \right\}$. 
The equation (\ref{eq:ssf}) indeed yields  
\begin{eqnarray*}
f_6 & = & 2 \left( H^{(2)}_{2} \right)^2 + H^{(1)}_{2} H^{(1)}_{3} \\
& = &
2 \left|
\begin{array}{cc}
a_2 & a_3 \\
a_3 & a_4 
\end{array}
\right|^2
+
\left|
\begin{array}{cc}
a_1 & a_2 \\
a_2 & a_3 
\end{array}
\right| \cdot 
\left|
\begin{array}{ccc}
a_1 & a_2 & a_3 \\
a_2 & a_3 & a_4 \\
a_3 & a_4 & a_5 
\end{array}
\right| \\
& = &
2 \left|
\begin{array}{cc}
-5\sigma & 21 \\
21 & -94\sigma 
\end{array}
\right|^2
+
\left|
\begin{array}{cc}
1 & -5\sigma \\
-5\sigma & 21 
\end{array}
\right| \cdot 
\left|
\begin{array}{ccc}
1 & -5\sigma & 21 \\
-5\sigma & 21 & -94\sigma \\
21 & -94\sigma & 443  
\end{array}
\right| \\
& = & 2\cdot 29^2 + (-4)(-129) \\
& = & 2198 \\ & = & 7\cdot 314 \\
& = & s_6 s_9. 
\end{eqnarray*}

Next we verify (\ref{eq:sbyH}). 
Since the parameters that correspond to $\left\{ nR \right\}$ are
$a = \sigma$, $b = 1$, $c = \sigma$, and  we obtain the sequence as 
$a_0 = a = \sigma$, $a_1 = b = 1$, $a_2 = c = \sigma$, 
$a_3 = 2 a_0 a_2 + a_1^2 = 3$, $a_4 = 2(a_0 a_3 + a_1 a_2) = 8 \sigma$, $a_5=23$, $\cdots$, 
which is \href{http://oeis.org/A025262/}{A025262} in \cite{OEIS} in the case $\sigma = 1$. 
The equation (\ref{eq:sbyH}) yields  
\begin{eqnarray*}
s_6 & = &  
\left|
\begin{array}{cccc}
a_1 & a_2 & a_3 & a_4\\
a_2 & a_3 & a_4 & a_5 \\
a_3 & a_4 & a_5 & a_6 \\
a_4 & a_5 & a_6 & a_7 
\end{array}
\right| 
 = 
\left|
\begin{array}{cccc}
1 & \sigma & 3 & 8 \sigma \\
\sigma & 3 & 8 \sigma & 23 \\
3 & 8 \sigma & 23 & 68 \sigma \\
8 \sigma & 23 & 68 \sigma & 207 
\end{array}
\right| 
 = 7. 
\end{eqnarray*}

\end{Example}

\section{Concluding remarks}
\label{sec:remarks}

In this paper, we give the explicit formulae for the elliptic sequence by means of the Hankel determinants. 
The formulae are being expected to contribute to enumeration in combinatorics
  or algorithmic number theory through elliptic curves
  because determinants have linear algebraic structure behind them. 
As an application, the solution of Somos-(4) by Hankel determinants is shown
  through the elliptic sequence.
The prime appearing and co-primeness of the general Somos $4$ will be future problems.
Integrable aspects of combinatorics or number theory seem to be interesting future problems, 
for example, application of Toda and Painlev\'e equations in a similar manner will also be
  interesting\cite{Kajiwara2001}. 

\ack
The author would like to thank M. Kanki and T. Tokihiro for fruitful discussions
  about the Somos sequence and its integrarity,
  and M. Shirase for many interesting discussions about elliptic curves. 
The author would also like to thank P. Barry for valuable comments to the first version of 
  this paper.

\appendix

\section{Proof of theorem \ref{thm:main}}
\label{sec:proof}

In this section, we prove theorem \ref{thm:main}. 
In the case $0 \le n \le 2$, (\ref{eq:EDS2}) may be easily checked. 
For $n \geq 3$, 
  let us substitute $n=2k+1$ or $n=2(k+1)$ \ $(k \geq 1)$ into (\ref{eq:EDS2}), 
  depending on the parity of $n$:
\begin{eqnarray}
  H^{(1)}_{k-1} H^{(1)}_{k+1} + (c-2ab) H^{(2)}_{k-1} H^{(2)}_{k} - b \left(H^{(1)}_{k} \right)^2 = 0 , 
    \label{eq:TBS2r} \\
  W_2^2 \left[ H^{(2)}_{k-1} H^{(2)}_{k+1} + H^{(1)}_{k} H^{(1)}_{k+1} - b \left(H^{(2)}_{k} \right)^2 \right] = 0 . 
    \label{eq:TBS1r}
\end{eqnarray}
If $W_2=0$, (\ref{eq:TBS1r}) trivially holds and (\ref{eq:TBS2r}) reduces ``Somos 2'':
\begin{equation}
  H^{(1)}_{k-1} H^{(1)}_{k+1} = b \left(H^{(1)}_{k} \right)^2, 
    \label{eq:TBS2rr}
\end{equation}
due to $c=2ab$. This ``Somos 2'' may be solved as $H^{(1)}_{n} = b^{n(n+1)/2}$ 
  (cf. Example \ref{ex:Catalan}).
This solution therefore reproduces 
  $W_{2n+1} = (-1)^n b^{n(n+1)/2}$ and $W_{2n} = 0$ for $n\geq 0$, 
  which was shown in \cite[Thm. 23.1]{Ward1948}. 

Hereafter we assume $W_2 \neq 0$. Then 
(\ref{eq:EDS2}) is equivalent to the following two equations with the definition (\ref{eq:defW}); 
\begin{eqnarray}
  H^{(1)}_{k-1} H^{(1)}_{k+1} + (c-2ab) H^{(2)}_{k-1} H^{(2)}_{k} - b \left(H^{(1)}_{k} \right)^2 = 0 , 
    \label{eq:TBS2} \\
  H^{(2)}_{k-1} H^{(2)}_{k+1} + H^{(1)}_{k} H^{(1)}_{k+1} - b \left(H^{(2)}_{k} \right)^2 = 0 . 
    \label{eq:TBS1}
\end{eqnarray}
The proof of these equations is similar to \cite{Chang2012}. 
We first prepare the several notations; 
\begin{eqnarray}
B_2 := a_2, B_k := a_k - \sum_{i=2}^{k-1} a_i B_{k-i+1}/a_1\ (k \geq 3), \\
L^{(m)}_{0} := 0, L^{(m)}_{1} := B_2 = a_2,
  \label{eq:convL} \\
L^{(m)}_{n} :=  
\begin{array}{|ccccc|}
B_{2} &  a_{m}   & a_{m+1} & \cdots & a_{m+n-2} \\
B_{3} &  a_{m+1} & a_{m+2} & \cdots & a_{m+n-1} \\
\vdots & \vdots & \vdots & & \vdots \\
B_{n+1} &  a_{m+n-1} & a_{m+n} & \cdots & a_{m+2n-3}  
\end{array} \quad (n \geq 2), 
  \label{eq:Lmn} \\
M^{(m)}_{1} := B_3,
  \label{eq:convM} \\
M^{(m)}_{n} :=  
\begin{array}{|ccccc|}
B_{3} &  a_{m}   & a_{m+1} & \cdots & a_{m+n-2} \\
B_{4} &  a_{m+1} & a_{m+2} & \cdots & a_{m+n-1} \\
\vdots & \vdots & \vdots & & \vdots \\
B_{n+2} &  a_{m+n-1} & a_{m+n} & \cdots & a_{m+2n-3}  
\end{array} \quad (n \geq 2). 
  \label{eq:Mmn}
\end{eqnarray}
Note also that $B_k$ is the invert transform of $a_k$\cite{Layman2001}. 

\begin{Proposition}
For $n \geq 2$, 
\begin{equation}
H^{(1)}_{n} =\left\{
  \begin{array}{ll}
  b^{n-2} \left[ b^2 H^{(1)}_{n-1} + (2ab-c) L^{(2)}_{n-1} \right] & (b \neq 0) \\
  -c^n H^{(2)}_{n-2} & ( b = 0 )
  \end{array}
\right. .
  \label{eq:136}
\end{equation}
\end{Proposition}
\begin{Proof}
In the case $n=2$, (\ref{eq:136}) follows from direct calculations
  under the convention (\ref{eq:convL}). 
For $n\geq 3$, subtracting 
  $\sum_{i=1}^{n-2} (\mbox{$i$th column}) \times a_{n-1-i}$
  and $(\mbox{$(n-1)$st column}) \times 2a_{0}$ 
  from $n$th column of $H_n^{(1)}$, we obtain 
\begin{eqnarray*}
H_n^{(1)}
  & = & 
\begin{array}{|cccc|}
a_{1} & \cdots & a_{n-1} & a_{n  } \\
a_{2} & \cdots & a_{n  } & a_{n+1} \\
\vdots & \vdots & \vdots & \vdots \\
a_{n} & \cdots & a_{2n-2} & a_{2n-1}  
\end{array} \\
  & = & 
\begin{array}{|cccc|}
a_{1} & \cdots & a_{n-1} & 0 \\
a_{2} & \cdots & a_{n  } & a_{1} a_{n-1} \\
\vdots & \vdots & \vdots & \vdots \\
a_{n} & \cdots & a_{2n-2} & \sum_{i=1}^{n-1} a_{i} a_{2n-2-i}  
\end{array} . 
\end{eqnarray*}
By similar elementary column additions from $(n-1)$st to second column, we obtain
\begin{eqnarray}
H_n^{(1)}
  & = & 
\begin{array}{|ccccc|}
a_{1} & a_{2} - 2a_{0}a_{1} & 0 & \cdots & 0 \\
a_{2} & a_{1}a_{1}             & a_{1}a_{2} & \cdots & a_{1} a_{n-1} \\
a_{3} & a_{1}a_{2}+a_{2}a_{1} & a_{1}a_{3}+a_{2}a_{2} & \cdots & a_{1}a_{n}+a_{2}a_{n-1} \\
\vdots & \vdots & \vdots &   & \vdots \\
a_{n} & \sum_{i=1}^{n-1} a_{i} a_{n-i} &
   \sum_{i=1}^{n-1} a_{i} a_{n+1-i} & \cdots & \sum_{i=1}^{n-1} a_{i} a_{2n-2-i}  
\end{array} .
  \label{eq:H1c}
\end{eqnarray}
Next, let us consider the cofactor expansion along the first row.
The $(1, 1)$ minor of (\ref{eq:H1c}) leads to $a_1^{n-1} H^{(1)}_{n-1}$
  due to the row additions from above to bottom. 
The $(1, 2)$ minor of (\ref{eq:H1c}) equals
\begin{eqnarray}
\begin{array}{|cccc|}
a_{2} &  a_{1}a_{2} & \cdots & a_{1} a_{n-1} \\
a_{3} &  a_{1}a_{3}+a_{2}a_{2} & \cdots & a_{1}a_{n}+a_{2}a_{n-1} \\
\vdots & \vdots &   & \vdots \\
a_{n} & \sum_{i=1}^{n-1} a_{i} a_{n+1-i} & \cdots & \sum_{i=1}^{n-1} a_{i} a_{2n-2-i}  
\end{array}. 
  \label{eq:H1c12}
\end{eqnarray}
In the case $a_1=0$, (\ref{eq:H1c12}) yields $a_2^{n-1} H^{(2)}_{n-2}$
  by the row additions. 
Otherwise $a_1 \neq 0$, (\ref{eq:H1c12}) is as follows: 
  By subtracting the first row multiplied by $(a_2/a_1)$ from the second row, 
  (\ref{eq:H1c12}) turns into 
\begin{eqnarray}
\begin{array}{|cccc|}
a_{2} &  a_{1}a_{2} & \cdots & a_{1} a_{n-1} \\
a_{3}-a_{2}^2/a_{1} &  a_{1}a_{3} & \cdots & a_{1}a_{n} \\
\vdots & \vdots &   & \vdots \\
a_{n} & \sum_{i=1}^{n-1} a_{i} a_{n+1-i} & \cdots & \sum_{i=1}^{n-1} a_{i} a_{2n-2-i}  
\end{array}, 
  \label{eq:H1c12n}
\end{eqnarray}
and by repeating the similar row additions from second to $n$th row, we obtain
\begin{eqnarray}
a_1^{n-2} 
\begin{array}{|ccccc|}
B_{2} &  a_{2} & a_{2} & \cdots & a_{n-1} \\
B_{3} &  a_{3} & a_{4} & \cdots & a_{n} \\
\vdots & \vdots & \vdots & & \vdots \\
B_{n} &  a_{n} & a_{n+1} & \cdots & a_{2n-3}  
\end{array} 
  = a_1^{n-2} L^{(2)}_{n-1} . 
  \label{eq:H1c12B}
\end{eqnarray}
Combining these results and replacing $a_0 = a, a_1=b, a_2=c$,
  we obtain (\ref{eq:136}). 
\qed
\end{Proof}

\begin{Proposition}
For $n \geq 2$, 
\begin{equation}
H^{(1)}_{n} =\left\{
  \begin{array}{ll}
  b^{n-1} M^{(2)}_{n-1} & (b \neq 0) \\
  -c^n H^{(2)}_{n-2} & ( b = 0 )
  \end{array}
\right. .
  \label{eq:245}
\end{equation}
\end{Proposition}
\begin{Proof}
In the case $n=2$, (\ref{eq:245}) follows from direct calculations
  under the convention (\ref{eq:convM}).  
For $n\geq 3$, subtracting 
  $\sum_{i=2}^{n-2} (\mbox{$i$th column}) \times a_{n-1-i}$
  and $(\mbox{$(n-1)$st column}) \times 2a_{0}$ 
  from $n$th column of $H_n^{(1)}$, we obtain 
\begin{eqnarray*}
H_n^{(1)}
  & = & 
\begin{array}{|cccc|}
a_{1} & \cdots & a_{n-1} & a_{n  } \\
a_{2} & \cdots & a_{n  } & a_{n+1} \\
\vdots &  & \vdots & \vdots \\
a_{n} & \cdots & a_{2n-2} & a_{2n-1}  
\end{array} \\
  & = & 
\begin{array}{|cccc|}
a_{1} & \cdots & a_{n-1} & a_{1} a_{n-2} \\
a_{2} & \cdots & a_{n  } & a_{1} a_{n-1} + a_{2} a_{n-2} \\
\vdots &  & \vdots & \vdots \\
a_{n} & \cdots & a_{2n-2} & \sum_{i=1}^{n} a_{i} a_{2n-2-i}  
\end{array}. 
\end{eqnarray*}
By similar method in the previous proof, we obtain
\begin{equation}
\hspace*{-20mm} 
H_n^{(1)} =  
\begin{array}{|cccccc|}
a_{1} & a_{2} & a_{1} a_{1}  & a_{1} a_{2}  & \cdots & a_{1} a_{n-2} \\
a_{2} & a_{3} & a_{1} a_{2}+a_{2} a_{1}  & a_{1} a_{3} + a_{2} a_{2}  & \cdots & a_{1} a_{n-1}+a_{2} a_{n-2} \\
\vdots & \vdots & \vdots & & & \vdots \\
a_{n} & a_{n+1} & \sum_{i=1}^{n} a_{i} a_{n+1-i} & \sum_{i=1}^{n} a_{i} a_{n+2-i} & \cdots & \sum_{i=1}^{n} a_{i} a_{2n-2-i}  
\end{array} .
  \label{eq:233}
\end{equation}
In the case $a_1=0$, the cofactor expansion along the first row gives $H^{(1)}_{n} = -a_2^n H^{(2)}_{n-2}$.
Otherwise $a_1 \neq 0$, (\ref{eq:233}) is as follows: 
  By subtracting the first row multiplied by $(a_2/a_1)$ from the second row
  and repeating the similar row additions from above to bottom, 
  (\ref{eq:233}) turns into 
\begin{eqnarray*}
\hspace*{-20mm} 
H_n^{(1)}
  & = & 
\begin{array}{|cccccc|}
a_{1} & a_{2}                   & a_{1} a_{1}  & a_{1} a_{2}  & \cdots & a_{1} a_{n-2} \\
0 & a_{3}-a_{2}^2/a_{1} & a_{1} a_{2}  & a_{1} a_{3}  & \cdots & a_{1} a_{n-1} \\
\vdots & \vdots & \vdots & & & \vdots \\
a_{n} & a_{n+1} & \sum_{i=1}^{n} a_{i} a_{n+1-i} & \sum_{i=1}^{n} a_{i} a_{n+2-i} & \cdots & \sum_{i=1}^{n} a_{i} a_{2n-2-i}  
\end{array} 
  \label{eq:234} \\
\hspace*{-20mm} 
& \ \vdots & \\ 
\hspace*{-20mm} 
& = & 
\begin{array}{|cccccc|}
a_{1} & a_{2}                   & a_{1} a_{1}  & a_{1} a_{2}  & \cdots & a_{1} a_{n-2} \\
0 & B_3 & a_{1} a_{2}  & a_{1} a_{3}  & \cdots & a_{1} a_{n-1} \\
\vdots & \vdots & \vdots & & & \vdots \\
0    & B_{n+1} & a_{1} a_{n} & a_{1} a_{n+1} & \cdots & a_{1} a_{2n-3}  
\end{array} 
  \label{eq:235} \\
\hspace*{-20mm} 
& = & a_1^{n-1} M^{(2)}_{n-1} . 
\end{eqnarray*}
This ends the proof of  (\ref{eq:245}). 
\qed
\end{Proof}

\begin{Proposition}
For $n \geq 1$, 
\begin{equation}
H^{(2)}_{n} =\left\{
  \begin{array}{ll}
  b^{n-1} L^{(1)}_{n} & (b \neq 0) \\
  c^n H^{(1)}_{n-1} & ( b = 0 )
  \end{array}
\right. .
  \label{eq:163}
\end{equation}
\end{Proposition}
\begin{Proof}
In the case $n=1$, (\ref{eq:163}) follows from direct calculations
  under the convention (\ref{eq:convL})
  as 
\begin{equation*}
H^{(2)}_{1} =\left\{
  \begin{array}{ll}
  b^{0} L^{(1)}_{1} = B_2 = c & (b \neq 0) \\
  c^1 H^{(1)}_{0} = c & ( b = 0 )
  \end{array}
\right. .
\end{equation*}
For $n\geq 2$, subtracting 
  $\sum_{i=1}^{n-2} (\mbox{$i$th column}) \times a_{n-1-i}$
  and $(\mbox{$(n-1)$st column}) \times 2a_{0}$ 
  from $n$th column of $H_n^{(2)}$, we obtain 
\begin{eqnarray*}
H_n^{(2)}
  & = & 
\begin{array}{|cccc|}
a_{2} & \cdots & a_{n  } & a_{n+1} \\
a_{3} & \cdots & a_{n+1 } & a_{n+2} \\
\vdots & \vdots & \vdots & \vdots \\
a_{n+1} & \cdots & a_{2n-1} & a_{2n}  
\end{array} \\
  & = & 
\begin{array}{|cccc|}
a_{2} & \cdots & a_{n   } & a_{1} a_{n-1} \\
a_{3} & \cdots & a_{n+1} & a_{1} a_{n  } + a_{2} a_{n-1} \\
\vdots & \vdots & \vdots & \vdots \\
a_{n+1} & \cdots & a_{2n-1} & \sum_{i=1}^{n} a_{i} a_{2n-1-i}  
\end{array} . 
\end{eqnarray*}
By similar elementary column additions from $(n-1)$st to second column, we obtain
\begin{eqnarray*}
H_n^{(2)}
  & = & 
\begin{array}{|ccccc|}
a_{2}   & a_{1} a_{1}              & a_{1} a_{2} & \cdots & a_{1} a_{n-1} \\
a_{3}   & a_{1} a_{2}+a_{2}a_{1}             & a_{1} a_{3}+a_{2}a_{2} & \cdots & a_{1} a_{n} + a_{2} a_{n-1} \\
\vdots & \vdots & \vdots &   & \vdots \\
a_{n+1} & \sum_{i=1}^{n} a_{i} a_{n+1-i} &
   \sum_{i=1}^{n} a_{i} a_{n+2-i} & \cdots & \sum_{i=1}^{n} a_{i} a_{2n-1-i}  
\end{array} .
\end{eqnarray*}
In the case $a_1=0$, the cofactor expansion along the first row yields $a_2^{n} H^{(1)}_{n-1}$. 
Otherwise $a_1 \neq 0$, by row additions similar to (\ref{eq:H1c12n}),
  we obtain $a_1^{n-1} L^{(1)}_n$. 
Thus, by $a_0 = a, a_1=b, a_2=c$, we obtain (\ref{eq:163}). 
\qed
\end{Proof}

\begin{Proof}[Equation (\ref{eq:TBS2}) and (\ref{eq:TBS1})]
We first prove the case $b=0$. In this case, (\ref{eq:136}) and (\ref{eq:163}) reduce to
\begin{eqnarray*}
H^{(1)}_{n} = -c^n H^{(2)}_{n-2} \ (n \geq 2), \quad H^{(1)}_{1} = b = 0, \quad H^{(1)}_{0} = 1, \\
H^{(2)}_{n} = c^n H^{(1)}_{n-1} \ (n \geq 1), \quad H^{(2)}_{0}= 1. 
\end{eqnarray*}
Then (\ref{eq:TBS2}) and (\ref{eq:TBS1}) hold as follows: 
\begin{eqnarray*}
  H^{(1)}_{k-1} H^{(1)}_{k+1} + (c-2ab) H^{(2)}_{k-1} H^{(2)}_{k} - b \left(H^{(1)}_{k} \right)^2
  \hspace{-60mm} 
  \\
& = &
  H^{(1)}_{k-1} H^{(1)}_{k+1} + c H^{(2)}_{k-1} H^{(2)}_{k} \\
& = & 0 , \\
  H^{(2)}_{k-1} H^{(2)}_{k+1} + H^{(1)}_{k} H^{(1)}_{k+1} - b \left(H^{(2)}_{k} \right)^2
  \hspace{-60mm} 
  \\
& = &
  H^{(2)}_{k-1} H^{(2)}_{k+1} + H^{(1)}_{k} H^{(1)}_{k+1} \\
& = & 0 . 
\end{eqnarray*}
Next, we consider the case $b \neq 0$. 
Since $W_2 \neq 0$ is assumed, we obtain $2ab-c = \sigma W_2^4 \neq 0$ and
\begin{eqnarray}
L^{(1)}_{n} & = & \frac{1}{b^{n-1}} H^{(2)}_{n}, \label{eq:L1} \\
L^{(2)}_{n-1} & = & \frac{1}{2ab-c} \left( \frac{1}{b^{n-2}} H^{(1)}_{n} - b^2 H^{(1)}_{n-1} \right), 
  \label{eq:L2} \\
M^{(2)}_{n-1} & = & \frac{1}{b^{n-1}} H^{(1)}_{n} \label{eq:M2}, 
\end{eqnarray}
from (\ref{eq:163}), (\ref{eq:136}) and (\ref{eq:245}), respectively.
The Jacobi identity for determinant 
\begin{equation}
A \left[ \begin{array}{cc} i & j \\ k & l \end{array} \right] A =
A \left[ \begin{array}{c} i \\ k \end{array} \right]
A \left[ \begin{array}{c} j \\ l \end{array} \right]
- 
A \left[ \begin{array}{c} i \\ l \end{array} \right]
A \left[ \begin{array}{c} j \\ k \end{array} \right] , 
  \label{eq:Jacobi}
\end{equation}
where 
$A \left[ \begin{array}{ccc} i_1 & \cdots & i_n \\ j_1 & \cdots & j_n \end{array} \right] $
  denotes the minor of $A$ without $i_1, \cdots, i_n$-th rows and $j_1, \cdots, j_n$-th column, is well-known. 
Applying the following $n \times n$ matrix
\begin{equation*}
A =  
\begin{array}{|ccccc|}
0 & 1 & 0 & \cdots & 0 \\
B_{2} & a_{m  } & a_{m+1} & \cdots & a_{m+n-2} \\
B_{3} & a_{m+1} & a_{m+2} & \cdots & a_{m+n-1  } \\
\vdots & \vdots & \vdots & & \vdots \\
B_{n} & a_{m+n-2} & a_{m+n-1} & \cdots & a_{m+2n-4}  
\end{array}
\end{equation*}
to (\ref{eq:Jacobi}) with $i=k=1$, $j=l=n$ yields
\begin{equation}
L^{(m+1)}_n H^{(m)}_{n-1} = L^{(m+1)}_{n-1} H^{(m)}_{n} + L^{(m)}_{n} H^{(m+1)}_{n-1}. 
\label{eq:173}
\end{equation}
Substituting  (\ref{eq:L1}) and (\ref{eq:L2}) into (\ref{eq:173}) with $m=1$, we obtain (\ref{eq:TBS2}).
Applying 
\begin{equation*}
A =  L^{(m)}_n = 
\begin{array}{|ccccc|}
B_{2} & a_{m  } & a_{m+1} & \cdots & a_{m+n-2} \\
B_{3} & a_{m+1} & a_{m+2} & \cdots & a_{m+n-1} \\
\vdots & \vdots & \vdots & & \vdots \\
B_{n+1} & a_{m+n-1} & a_{m+n} & \cdots & a_{m+2n-3}  
\end{array}
\end{equation*}
to (\ref{eq:Jacobi}) with $i=k=1$, $j=l=n$ yields 
\begin{equation}
L^{(m)}_n H^{(m+1)}_{n-2} = H^{(m+1)}_{n-1} L^{(m)}_{n-1} - M^{(m+1)}_{n-1} H^{(m)}_{n-1}. 
\label{eq:251}
\end{equation}
Substituting (\ref{eq:L1}) and (\ref{eq:M2}) into (\ref{eq:251}) with $m=1$, we obtain (\ref{eq:TBS1}).
These complete the proof of the thorem. 
\qed
\end{Proof}


\end{document}